# Electronic transmission in Graphene suppressed by interlayer interference

Daniel Valencia[1], Jun-Qiang Lu[1,*], Jian Wu[2], Feng Liu[3], Feng Zhai[4], Yong-Jin Jiang[4]

[1]Department of Physics and Institute for Functional Nanomaterials, University of Puerto Rico, Mayaguez, PR 00681, USA

[2]Department of Physics, Tsinghua University, Beijing 100084, China

[3]Department of Materials Science and Engineering, University of Utah, Salt Lake City, Utah 84112, USA

[4]Department of Physics and Center for Statistical Physics and Condensed Matter Physics, Zhejiang Normal University, Jinhua, Zhejiang 321004, China

*Email: junqiang.lu@upr.edu

**ABSTRACT**

We investigate electronic transport property of a graphene monolayer covered by a graphene nanoribbon. We demonstrate that electronic transmission of a monolayer can be reduced when covered by a nanoribbon. The energy at which the transmission reduction occurs depends on the width of nanoribbon. We explain the transmission reduction as interference between wavefunctions in the monolayer and the nanoribbon. Furthermore, we show that the transmission reduction of a monolayer is *combinable* when covered by more than one nanoribbon and we propose a concept of "combination of control" for possible nano-application designs.

Graphene is the first atomic monolayer structure fabricated experimentally.[1] Its unique geometric structure (2-D) along with electronic structure (Dirac cone) makes it an ideal system for fundamental exploration of novel physics as well as a promising candidate for nanoelectronic applications.[2] Among many proposals to explore fundamental physics in graphene structures, the so-called Klein paradox or Klein tunneling has recently stimulated extensive research interests.[3-5] The Klein tunneling in graphene structures is a result of the chiral nature of the Dirac cone band structures, which is analogous to the chirality in three-dimensional quantum electrodynamics. It shows a local electrostatic barrier applied by a gate on top of a graphene monolayer is perfectly transparent for normal incident charge carriers in the monolayer. Though the Klein tunneling is theoretically interesting, it creates disadvantage in applications as the perfect transmission may limit the performance of graphene-based electronic devices like diodes and transistors.[6]

Though the perfect transmission of a graphene monolayer cannot be blocked by a gate voltage, in this paper, we demonstrate that it can be suppressed by simply replacing the gate voltage with a graphene nanoribbon. We illustrate our idea by calculating electronic transmission[7] of graphene mono-bi-monolayer junctions, as shown in Fig. 1. Our results show that though the transmission spectra of these junctions are similar to that of a monolayer, their transmission is reduced at certain energies. The transmission reduction at these energies is attributed to antiresonance due to interference between the wavefunctions in two layers. Moreover, we show that nanoribbons with different widths reduce transmission of the monolayer at different energies, thus the prefect transmission of a graphene monolayer can be further reduced if it is covered by more than one nanoribbon.

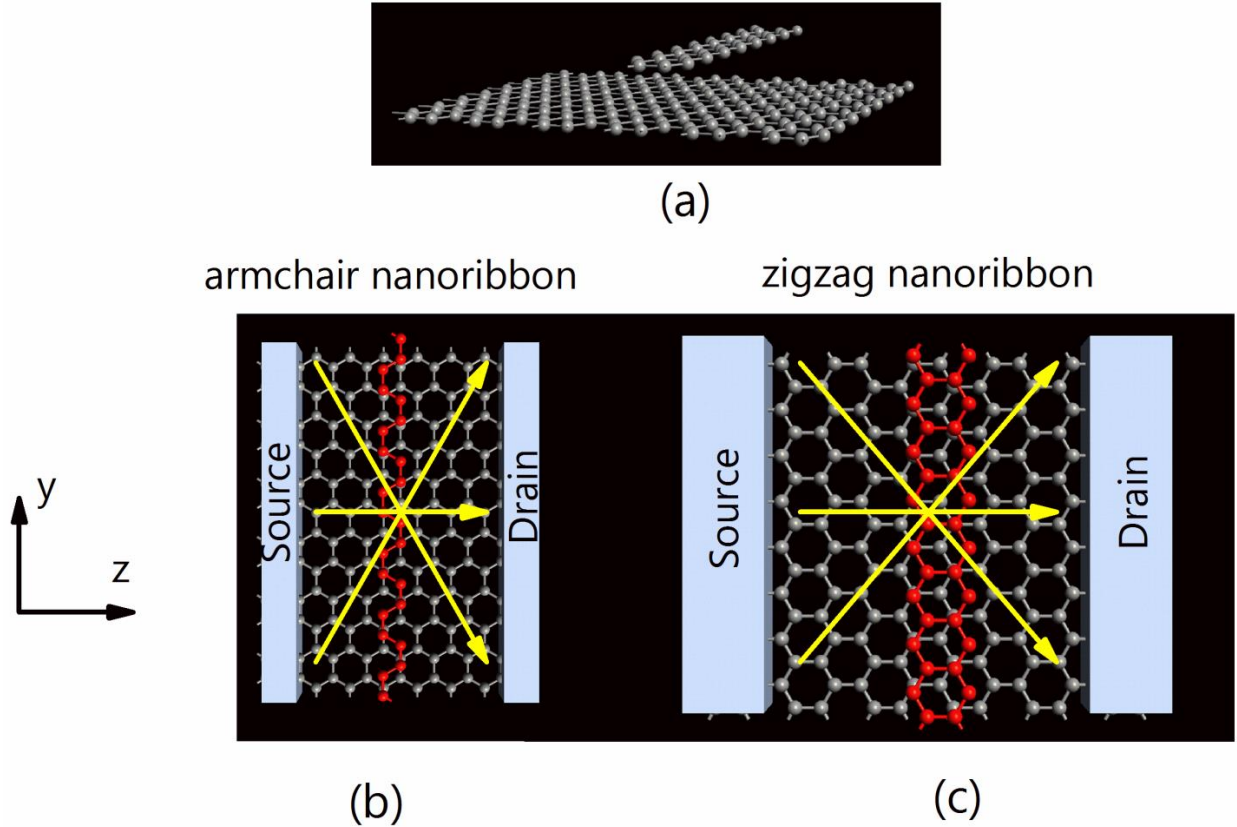

Fig. 1 (Color online) (a) A graphene mono-bi-monolayer junction, or a monolayer covered by a nanoribbon. (b) & (c) Schematic setups to study electronic transport properties of different junctions: a monolayer covered by (b) an armchair nanoribbon or (c) a zigzag nanoribbon. Electronic transmissions are calculated along different directions as shown by the yellow arrows in (b) and (c).

Fig. 1(a) presents the typical structure studied in this paper: a monolayer covered by a nanoribbon (or a graphene mono-bi-monolayer junction). The distance between the nanoribbon and the monolayer is 3.383 Å. The nanoribbon can be an armchair or zigzag one depending on its direction or edge. We deposit source and drain electrodes parallel to the direction of the nanoribbon as shown in Fig. 1(b) and 1(c), and calculate electronic transmission from source to drain through the monolayer covered by the nanoribbon (or through the junction). From the source to the drain, electrons can travel through the monolayer along different directions (or transmission channels), as shown by the arrows in Fig. 1(b) and 1(c). Different transmission channels can be identified by different wave vectors along the transverse direction ($k_y$). To calculate the transmission spectra of the junctions, we use the *ab initio* simulation package: Atomistix,[8] which calculates electronic transport properties of nanostructures based on nonequilibrium Green function and density functional theory.[9,10] In our calculations, we use the Double Zeta Polarized basic set, and the exchange-correlation potential of the generalized gradient approximation with the Perdew-Burke-Ernzerhof parameterization. The energy cutoff is 70.0 Ry, and the convergence threshold is $10^{-5}$ eV. All calculations use room temperature.

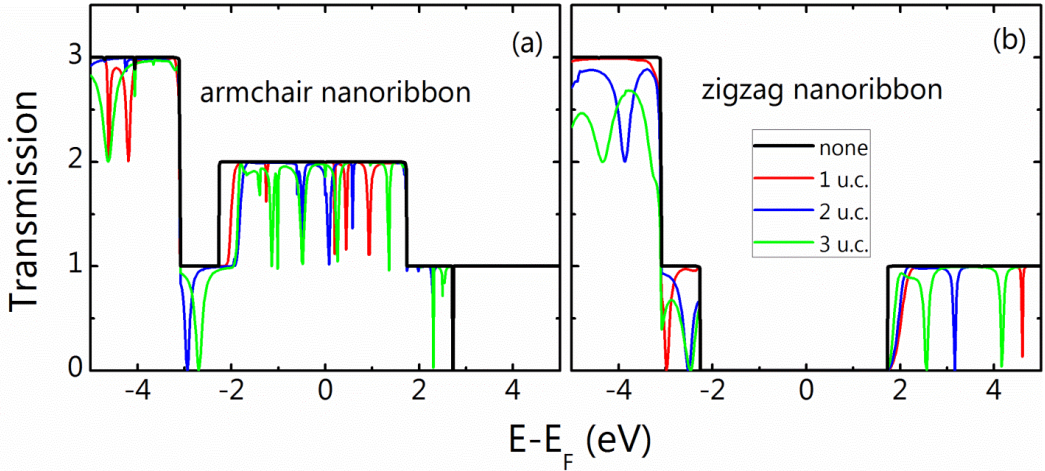

Fig. 2 (Color online) Normal incident transmission spectra of a monolayer covered by (a) an armchair nanoribbon or (b) a zigzag nanoribbon, with the nanoribbon width of one (red), two (blue), or three (green) unit cells. The transmission spectra (black) of a monolayer without coverage of nanoribbon are also plotted for reference.

Fig. 2 plots the normal incident ($k_y$ = 0) transmission spectra of a monolayer covered by a nanoribbon of varying widths. Fig. 2(a) is for the case covered by an armchair nanoribbon, as shown in Fig. 1(b), and Fig. 2(b) is for the case covered by a zigzag nanoribbon, as shown in Fig. 1(c). These results show that the transmission of a monolayer is reduced when covered by a nanoribbons. Moreover, the transmission is reduced to zero at certain energies. When the width of nanoribbon increases from one to three unit cells (u.c.), the number of zeros in transmission increases and their positions change.

The zeros in transmission can be understood by interference between electron wavefunctions in the monolayer and the nanoribbon. For a normal incident ($k_y$ = 0) electron with a particular energy E and wave vector $k_z$, it has one transmission channel in the monolayer before reaching the region covered by the nanoribbon. Once it reaches the covered region, the electron can travel through another channel available in the nanoribbon due to interlayer coupling. Thus the covered region provides two channels for electron transmission, as shown in Fig. 3(a). Since electron wavefunctions exist in both channels, destructive interference between them occurs at certain wave vectors ($k_z$). This is the so-called antiresonance,[11-13] which leads to the zeros in the transmission spectra.

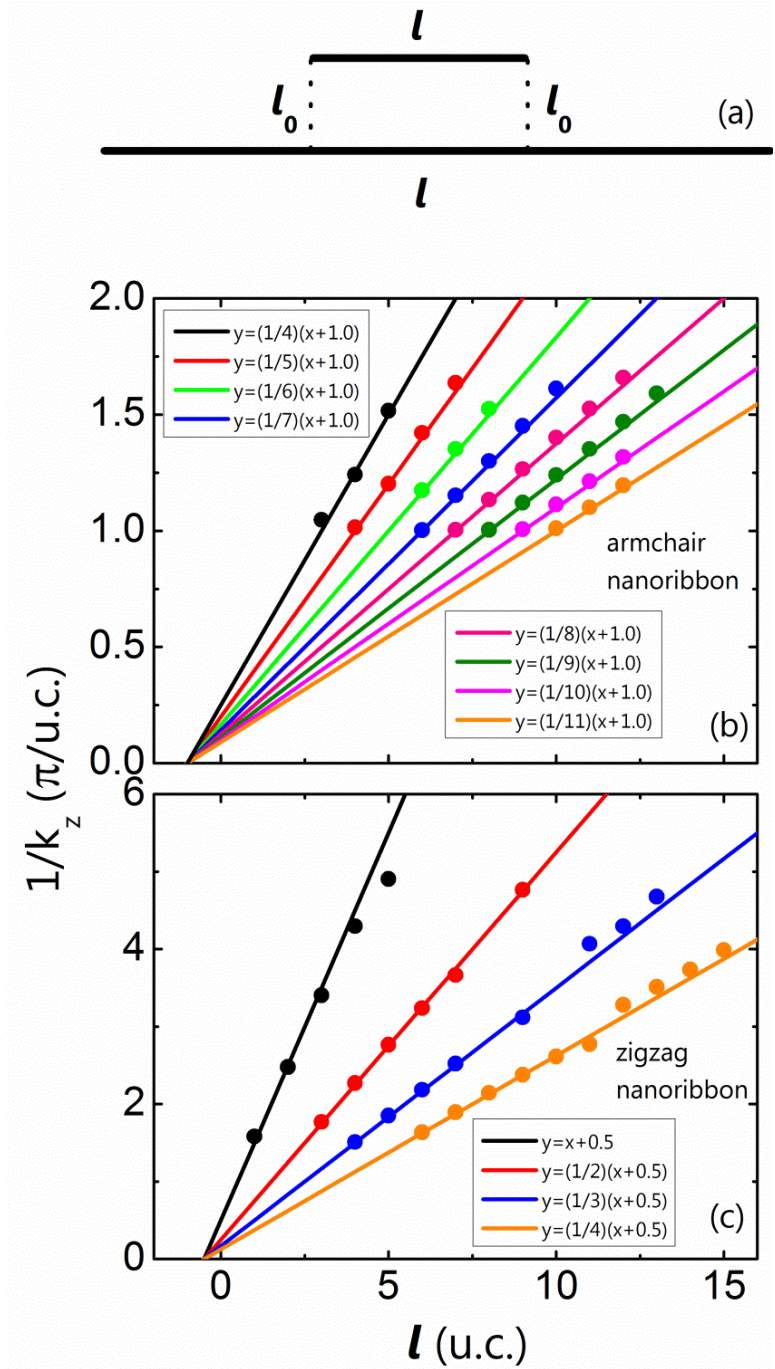

Fig. 3 (Color online) (a) A schematic diagram of transmission channels of a monolayer covered by a nanoribbon; $l$ is the width of nanoribbon and also the width of two-channel region; $l_o$ is an effective length used to represent interlayer hopping distance. (b) & (c) The inverse of the wave vectors $k_z$ at antiresonance as functions of the width of (b) an armchair nanoribbon or (c) a zigzag nanoribbon.

The antiresonance in transmission occurs if change of phase of wavefunction equals $2n\pi$ when the electron travels one loop of the closed path in the two-channel region, e.g. $k_z(2l+2l_0)=2n\pi$, where $l$ is the width of nanoribbon and $l_o$ is an effective length used to represent interlayer hopping distance, as shown in Fig. 3(a). Thus at antiresonance, the wave vector $k_z$ should be inversely proportional to the width of nanoribbon, e.g. $\frac{1}{k_z}=\frac{1}{n\pi}(l+l_0)$. In Fig. 3(b) and 3(c), we plot the inverse of wave vector $k_z$ at zero transmission as a function of width of nanoribbon $l$. Fig. 3(b) is for the case covered by an armchair nanoribbon, and Fig. 3(c) is for the case covered by a zigzag nanoribbon. Since the unit of $k_z$ is $\pi/\text{u.c.}$, and the unit of $l$ is u.c., both Fig. 3(b) and 3(c) show the relationship $\frac{1}{k_z}=\frac{1}{n}(l+l_0)$, which confirms the antiresonance mechanism.

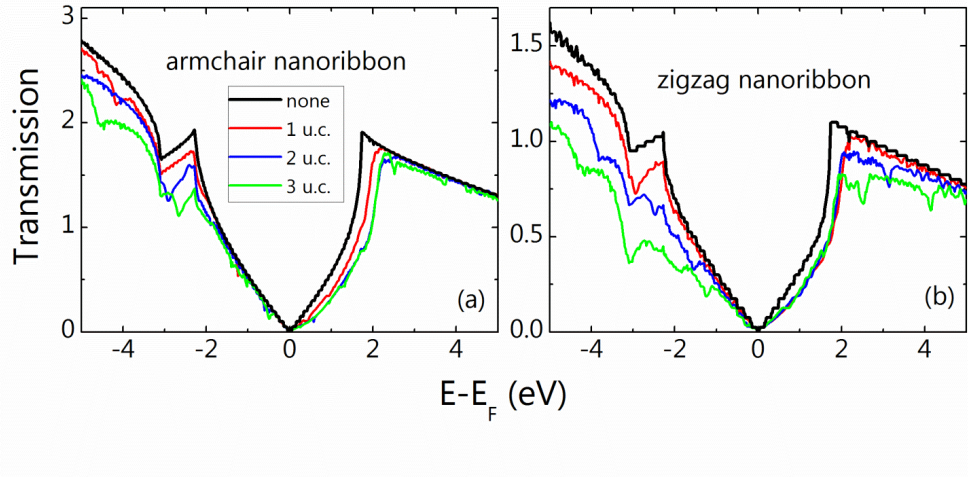

Fig. 4 (Color online) Total transmission (per unit cell) spectra of a monolayer covered by (a) an armchair nanoribbon or (b) a zigzag nanoribbon, with the width of one (red), two (blue), or three (green) unit cells. The transmission spectra (black) of a monolayer without coverage of nanoribbon are also plotted for reference.

The above results clearly show that a nanoribbon is not transparent to normal incident electrons in the monolayer: it can reduce electronic transmission at certain wave vectors or energies. However, reducing or changing transmission of normal incident electrons only is not enough for applications, since electrons can also go from source to drain through channels with other $k_y$s, as shown in Fig. 1(b) and 1(c). Fortunately, the linear relation and the antiresonance mechanism discussed above are $k_y$-independent, so they are general and can be applied to other $k_y$ channels. In other words, a nanoribbon placed on top of a monolayer will also reduce the transmission in other $k_y$ channels and hence the total transmission (summed up the transmission in all channels). To illustrate the idea, Fig. 4 presents the total transmission spectra (per unit cell) of a monolayer covered by a nanoribbon with different widths. Fig. 4(a) is for the case covered by an armchair nanoribbon, and Fig. 4(b) is for the case covered by a zigzag nanoribbon. In both Fig. 4(a) and (b), as the width of nanoribbon increases from one unit cell to four unit cells, the transmission of the monolayer decreases. Thus, electronic transmission of a monolayer can be reduced by depositing a nanoribbon of finite width on top of it.

In order to further reduce electronic transmission of a monolayer, we can simply deposit more than one nanoribbon on top of it. From the above discussion of antiresonance, a nanoribbon can only reduce transmission at certain wave vectors or energies determined by the width of the nanoribbon. Thus, two nanoribbons of different widths can reduce transmission of a monolayer at different wave vectors or energies when both of them are deposited on top of the monolayer, as shown in Fig. 5(a). Fig. 5(b) and 5(c) plot normal incident transmission spectra of a monolayer covered by two parallel nanoribbons of one and two unit cells. Fig. 5(b) is for the case of two armchair nanoribbons and Fig. 5(c) is for the case of two zigzag nanoribbons. In both Fig. 5(b) and 5(c), the normal incident transmission spectra of a monolayer covered by a one-unit-cell nanoribbon only or a two-unit-cell nanoribbon only are also plotted for reference. The transmission spectra are calculated for different spacing between the two parallel nanoribbons (1-1-2 in Fig. 5 for one-unit-cell spacing and 1-2-2 for two-unit-cell spacing); however, they are almost independent of the spacing: the transmission is reduced at the same energies for different spacing. More importantly, the transmission reduction by two nanoribbons together is the combination of the reduction by each nanoribbon separately. Thus the transmission of a monolayer can be reduced even further if more nanoribbons of varying widths are deposited on top of it.

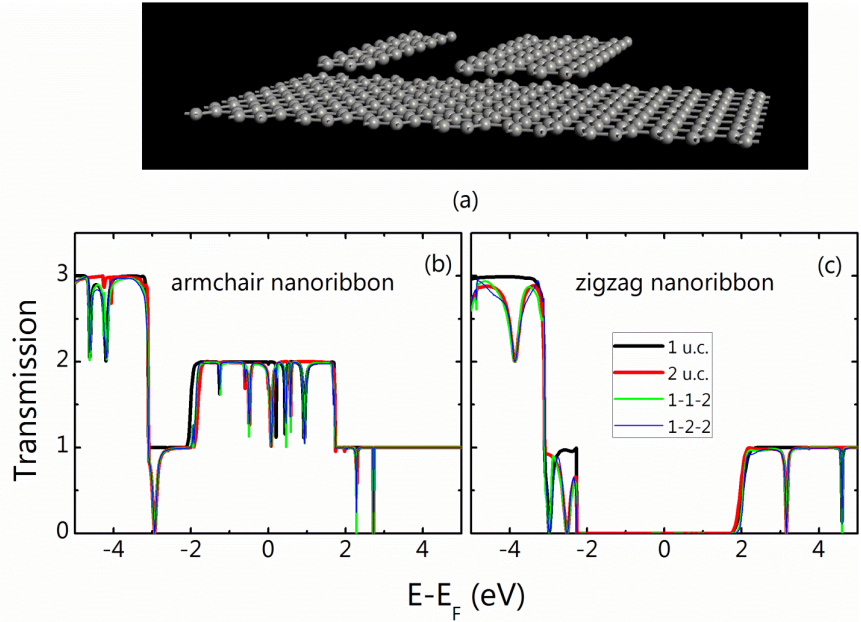

Fig. 5 (Color online) (a) A graphene monolayer covered by two nanoribbons of different widths. (b) & (c) Normal incident transmission spectra of a monolayer covered by (b) two armchair nanoribbons or (c) two zigzag nanoribbons, with the widths of one and two unit cells; the two nanoribbons are separated by one (green) or two (blue) unit cells. The transmission spectra of a monolayer covered by a single nanoribbon with width of one-unit-cell (black) or two-unit-cell (red) are also plotted for reference.

Different from a usual quantum tunneling junction[7] which *allows* transmission at certain energies, a nanoribbon *blocks* the transmission of a monolayer at certain energies. The unique *blocking* mechanism makes the combination of reduction possible and may lead to a concept of "combination of control" in application. Here we propose a conceptual design for a nano-keypad. As shown in Fig. 6, pushing down a combination of keys will lead to a corresponding combination of transmission reduction; in other words, a combination of keys controls a unique transmission spectrum.

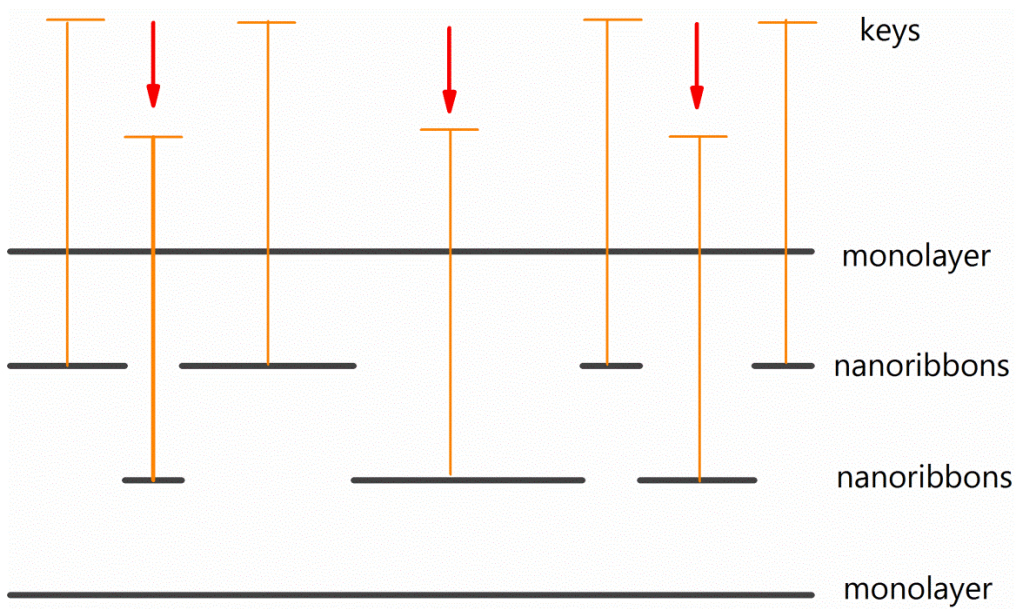

Fig. 6 (Color online) A schematic diagram of conceptual design of a nano-keypad based on the possibility of combination of transmission reduction by individual nanoribbon.

In summary, we demonstrate an approach to reduce electronic transmission of a graphene monolayer by covering it with a nanoribbon, instead of controlling it with a gate voltage. The underlying physics of transmission reduction is the interference between the wavefunctions in the monolayer and the nanoribbon. When covered by more than one nanoribbon, the transmission reduction of a monolayer is the combination of the reduction by each nanoribbon. Based on the combination of transmission reduction, we proposed a concept of "combination of control" and a possible design of a nano-keypad.

*This work is supported by NSF-EPSCOR program (Grants 1002410 and 1010094) and an award from Research Corporation for Science Advancement.*